\documentclass[12pt]{amsart}
\usepackage{amsmath,amsfonts,amssymb,amsthm,verbatim,multirow,url,subfig,footnote,graphicx,array,xr,booktabs,placeins,float,setspace}
\usepackage[usenames,dvipsnames]{color}
\usepackage[utf8]{inputenc}
\usepackage[T1]{fontenc}
\usepackage[version=3]{mhchem}
\usepackage[compress,authoryear]{natbib}
\usepackage[top=1in, bottom=1in, left=1in, right=1in]{geometry}

\theoremstyle{plain}

\theoremstyle{definition}

\theoremstyle{remark}

\begin{document}
\doublespacing
\title[]{Efficient Reconstructions of Common Era Climate via Integrated Nested Laplace Approximations.}

\author[]{Luis A. Barboza}
\address{Centro de Investigacion en Matematica Pura y Aplicada (CIMPA)-Escuela
  de Matematica, Universidad de Costa Rica\\
San Jos\'e, Costa Rica}
\email{luisalberto.barboza@ucr.ac.cr}

\author[]{Julien Emile-Geay}
\address{Department of Earth Sciences \\
  University of Southern California \\
  Los Angeles, California, USA.
}
\email{julieneg@usc.edu}

\author[]{Bo Li}
\address{Department of Statistics \\
  University of Illinois at Urbana-Champaign \\
  Champaign, Illinois, USA.
}
\email{libo@illinois.edu}

\author[]{Wan He}

\date{\today}
\subjclass[2010]{}
\maketitle

\begin{abstract}
Paleoclimate reconstruction on the Common Era (1-2000AD) provide critical context for recent warming trends. This work leverages integrated nested Laplace approximations (INLA) to conduct inference under a Bayesian hierarchical model using data from three sources: a state-of-the-art prox database (PAGES 2k), surface temperature observations (HadCRUT4), and latest estimates of external forcings. 
INLA's computational efficiency allows to explore several model formulations (with or without forcings, explicitly modeling internal variability or not), as well as five data reduction techniques. Two different validation exercises find a small impact of data reduction choices, but a large impact for model choice, with best results for the two models that incorporate external forcings. These models confirm that man-made greenhouse gas emissions are the largest contributor to temperature variability over the Common Era, followed by volcanic forcing. Solar effects are indistinguishable from zero. INLA provide an efficient way to estimate the posterior mean, comparable with the much costlier Monte Carlo Markov Chain procedure, but with wider uncertainty bounds. We recommend using it for exploration of model designs, but full MCMC solutions should be used for proper uncertainty quantification.  
\end{abstract}

\section{Introduction.}
\label{sec:intro}

Earth's climate presents a continuum of variability, with periodic and non-periodic fluctuations ranging from 1 to $10^8$ years \citep{pelletier_power_1998}. In particular, variability on scales of decades to centuries is of paramount importance for adaption and planning to anthropogenic climate change, yet is incompletely sampled by the relatively short historical record of wide-spread instrumental observations, going back to about 1850 CE \citep{AR5_chap5}. It is thus critical to reconstruct these variations from the paleoclimate record as quantitatively as possible.  A particular focus has been reconstructions of global or hemispheric temperature from high-resolution proxy observations \citep{Jones_Holocene09}.

Many methods have been developed to reconstruct past climates, particularly temperatures:  principal component regression
\citep{MBH98,luterbacher2004european}, regularized forms of the expectation-maximization algorithm \citep{Schneider2001,mann2007robust,JEG10a,Guillot_AOAS2015}, canonical
correlation analysis \citep{smerdon2010pseudoproxy,Wang_CP2014}, pairwise comparison
\citep{Hanhijarvi2013}, data assimilation \citep{Lee_CD08,Hakim2016}, and Bayesian hierarchical models (BHMs). 

BHMs offer several distinct advantages for paleoclimatic reconstruction. They can (i) treat different sources of uncertainty in a natural way, (ii) incorporate prior knowledge of the system in a logically-coherent manner, and (iii) allow for both inference and prediction \citep{Tingley_QSR2012}. 
Many studies have employed BHMs \citep[e.g.][hereafter, "B14"]{boli1, tingley2013_Ext,Barboza2014}); some have used space-state
schemes to linearly relate information that comes from paleoclimate observations together with information about external climatic forcings, resulting from 
well-mixed greenhouse gases, volcanic activity, and variations in solar output. Three problems arise in this
case: the need to reduce dimensionality, the complexity/realism of the model, and the execution time of the numerical procedure. 

B14 proposed a method that jointly models the variability of the
temperature series (as a latent process) as well as the variability of those
climatic and biological observations that serve as approximations of this process (``proxies'').
The authors found that long memory error terms are necessary in absence of
external forcing information within a BHM, but the
presence of external forcing information substantially improves the reconstruction. The scope of the study was restricted by the computational requirements of the Monte Carlo Markov Chain (MCMC) procedure, which limited the sensitivity analysis and forced severe levels of data reduction. 

In this article we extend the work of B14 by leveraging Integrated Nested Laplace
Approximations (INLA). Doing so lightens the computational burden, which allows to (a) explore new model designs inspired by the physics of the problem; (b) consider various choices for data reduction; and (c) take the non-stationary nature of the observational network into account.  In addition, this work makes use of the latest estimates of radiative forcing, as well as a state-of-the-art, open-access compilation of community-curated paleoclimate observations \citep{PAGES2kSD2017short}. This ensures that our calculations are using the best available data and are completely reproducible.  

The article is organized as follows: we start by describing the datasets (section 2), and the methodology (section 3). Results are then presented in section 4, discussed in Section 5, before concluding in section 6. 

\section{Datasets.}
\label{sec:data}

\subsection{Proxy data.}
Reconstructions of past climates rely on ``proxies'': indirect observations of climate, as recorded in borehole, coral, documentary, glacier
ice, lake and marine sediment, sclerosponge, speleothem and tree-ring archives
\citep{Jones_Holocene09}. The PAGES2k global multiproxy database is a community-driven effort to
synthesize all publicly-archived, temperature-sensitive proxy records of the
past 2,000 years \citep{PAGES2kSD2017short}. The most recent effort gathers 692 records from 648 locations, including all continental regions and major ocean basins. The records are from trees, ice, marine and lake sediments, corals, speleothems, and documentary evidence. They range in length from 50 to 2000 years, with a median of 547 years, while temporal resolution ranges from biweekly to centennial.
  The vast majority of records are annually-resolved, with minimal dating uncertainty. Here, the data used have been mapped to a standard normal using the method of \cite{vanAlbada2007}. The data are available in a standard format (LiPD) readable in R, Python and Matlab, to ensure reproducible workflows \citep{lipd_cp}.   

\begin{figure}
  \centering
  \includegraphics[scale=0.40]{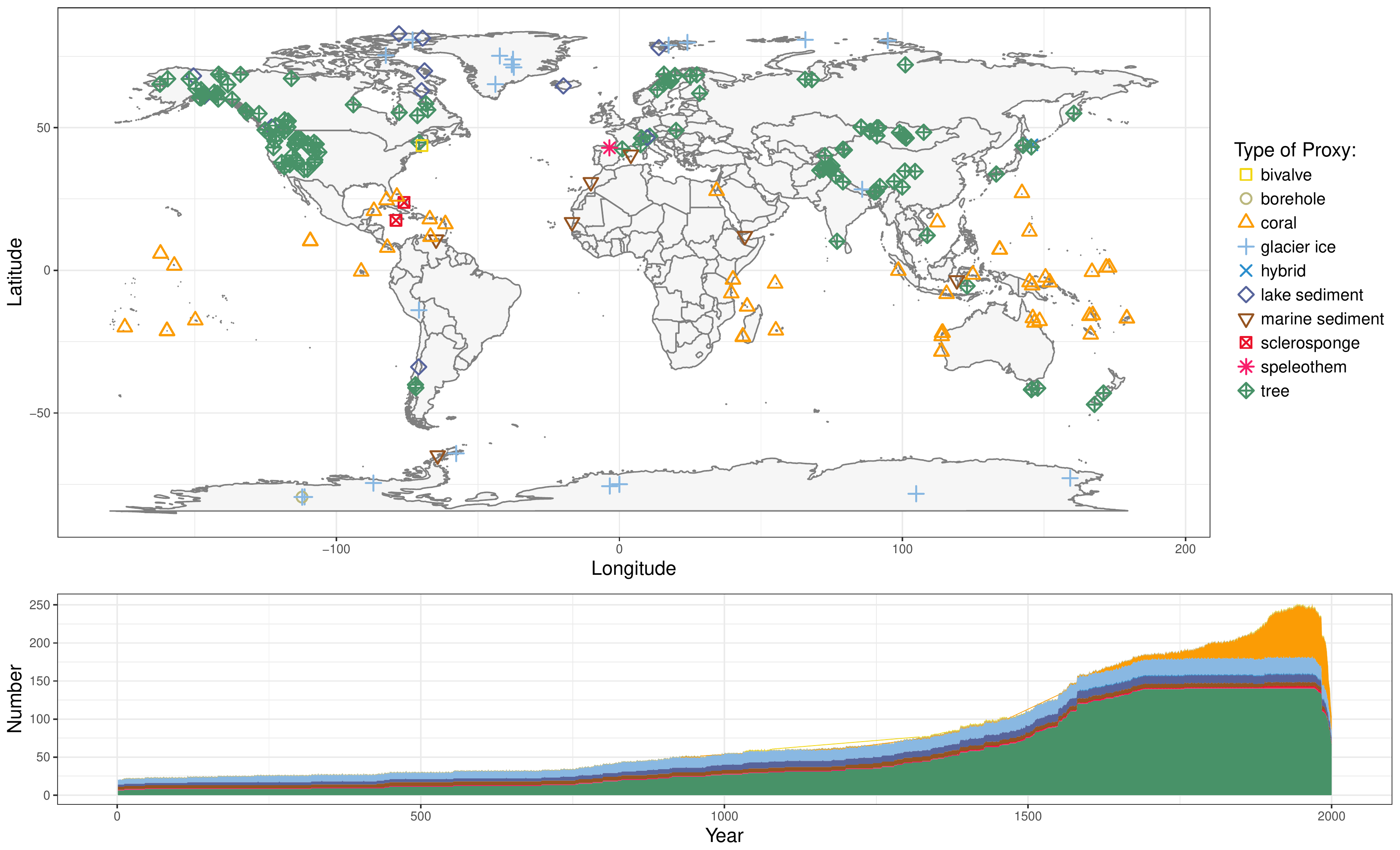}
  \caption{Distribution and temporal availability of PAGES2k proxies, after applying the screening procedure \cite{PAGES2kSD2017short}.}
  \label{fig:proxy}
\end{figure}

 Each of those proxies has different time horizons (Fig.~\ref{fig:proxy}, bottom), which creates challenges for inference. Unlike previous studies (e.g. B14), we strive to take into account the information available in most proxies, despite their temporal diversity.
In order to select proxies with high predictive power, we first chose those series with large correlations with respect to
their closest spatial temperature record in the HadCRUT4.2 dataset \citep{Morice2012}. More
details on this ``screening'' procedure, which controls for the multiple test problem \citep{BenjaminiHochberg95}, can be found in \citet{PAGES2kSD2017short}; it whittles down the database to 257 proxies (Fig.~\ref{fig:proxy}).   

\begin{table}[h!]
  \centering
  \begin{tabular}{c|c|c}
    \toprule
    Group & Interval (year AD) & Number of Proxies\\
    \midrule
    1 & 1-250 & 19 \\
    2 & 251-500 & 25 \\
    3 & 501-750 & 29 \\
    4 & 751-1000 & 33 \\
    5 & 1001-1250 & 54 \\
    6 & 1251-1500 & 65 \\
    7 & 1501-1750 & 105 \\
    8 & 1751-2000 & 146 \\
    \bottomrule
  \end{tabular}
  \caption{Distribution of proxies according to their temporal availability.}
  \label{tab:distdate}
\end{table}

\vskip -0.7cm

Due to the diversity of start dates in the proxies database (Fig.~\ref{fig:proxy}), we gather proxies into non-homogeneous groups where each group has temporal availability within a 250y interval (Table \ref{tab:distdate}).  Proxy series whose proportion of missing annual observations is larger than 5\% during the calibration period (1900-2000, as in  B14) are excluded.  

\subsection{Temperature data.}
We estimate Global Mean Surface Temperature (GMST) from the HadCRUT4 global temperature dataset provided by the Met Office Hadley
Centre and the Climatic Research Unit at the University of East Anglia, UK
(version 4.4.0.0). The dataset consists of instrumental, {\it in situ}
observations of surface temperature over land \citep{Jones2012} and ocean
\citep{Kennedy2011a,Kennedy2011}. The
observations are expressed as anomalies relative to the monthly-mean seasonal cycle over the 1961-1990 period, in degrees Celsius. Though HadCRUT4 features a rather sophisticated analysis of error sources \citep{Morice2012}, we neglect these uncertainties against the much larger uncertainties affecting paleoclimate observations, and simply use the median estimate, averaged on an annual basis.


\subsection{Forcing data.}
\begin{figure}
  \centering
  \includegraphics[scale=0.40]{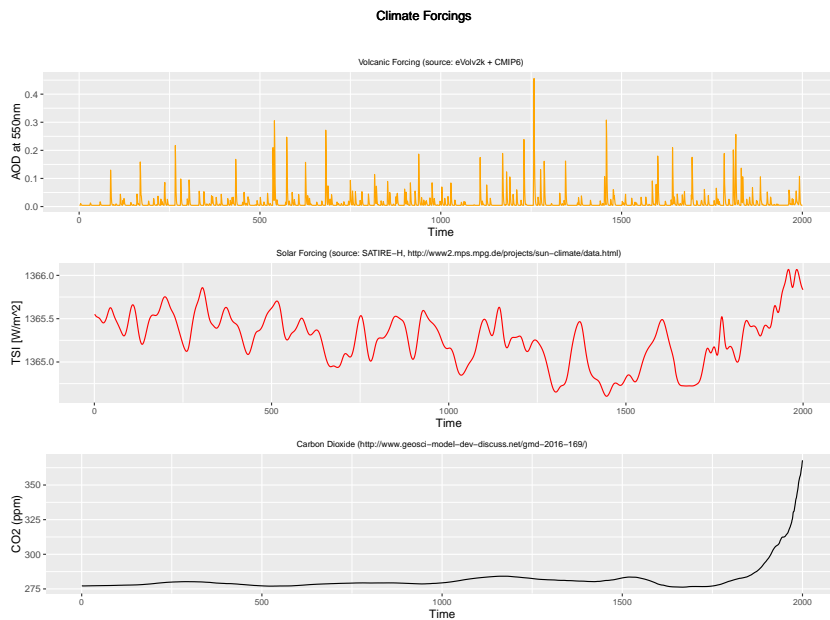}
  \caption{Main climate Forcings of the Common Era (1-2000 AD): volcanic, solar, and carbon dioxide. AOD = Aerosol optical depth. TSI = total solar irradiance. ppm = parts per million. See text for details.}
  \label{fig:forcings}
\end{figure}

These use the most recent compilations from the PMIP4-CMIP6 project \citep{JungclausGMD17}: 

\noindent {\bf Volcanic forcing} from the evolv2k dataset
  \citep{Toohey2016}, which reconstructed zonal mean aerosol optical depth at 550 nm, covering the 500 BCE to 1900 CE time period. For 1900 (or 1850) to present, \cite{Thomason2016} is
  used to fill in the forcing table. The data were given as a function of latitude, so were area-weighted and averaged to form a global estimate (Fig.~\ref{fig:forcings}, top).
  
\noindent {\bf Solar forcing} data is computed from the SATIRE-H
   dataset \citep{Vieira2011}. Irradiance is
  provided on a decadal basis from 9495BC - 1939AD and then on a daily basis
  from 1940AD onwards. We interpolated the data at annual resolution using BSplines (Fig.~\ref{fig:forcings}, middle). 
  
\noindent {\bf Greenhouse-Gases concentrations}: hemispheric means of mole
  fraction of carbon dioxide in air (ppm) with annual
  resolution \citep{Meinshausen_GMD2017} as well as ice core measurements prior to that date \citep{JungclausGMD17} (Fig.~\ref{fig:forcings}, bottom).

\section{Methodology}\label{sec:model}
\subsection{Data reduction methods}
\label{sec:rp}

For some intervals in Table \ref{tab:distdate}, the proxy data matrix is large compared to the 101 yearly samples used to train the model (1900-2000). This ``large $p$, small $n$'' problem calls for some form of data reduction. Following B14, we generate a set of ``Reduced Proxy'' ($RP$) time series, which condense individual proxy time series into a single time series with larger predictive power over the GMST target. Since this is a critical choice to make, we extend B14 by carefully investigating five common data reduction methods:



{\bf Lasso Regression (LR)}
  The Lasso regression penalizes the usual sum of squares with an argument
 containing the sum of the absolute values of each coefficient in the classical
 linear regression model, multiplied by an additional smoothing parameter \citep{Tibshirani1996}. Due
 to the geometric nature of the term of penalization, the search of estimators
 tends to assign values very close to zero to variables that have almost null
 effects with respect to the dependent variable, which makes the resulting
 models easily interpretable. This method is very common to data reduction and easy to implement, but it often tends to select overly complex models, that is,
 it tends to show "false positives" in the variable selection process 
 \citep{Fan2010}. The Lasso may also run into inconsistency issues when the
 variables are highly correlated \citep{Zou2005}.
 We used 10-fold cross validation to select the smoothing parameter when we carried out Lasso regression (see \cite{Tibshirani1996} for more details). 
 
{\bf Sparse Partial Least Squares (sPLS)}
  Partial least squares seek to reduce the high-dimensionality issues of the
  design matrix in  
  linear regression models through a latent matrix whose columns maximize
  the product of the linear correlation between predictors and responses and the
  variance of responses. The sPLS method further introduces sparseness to the partial least squares
  estimators by means of a $L_1$-penalty with a thresholding parameter, in order to avoid inconsistencies when there is a
  substantial number of noisy covariates \citep{Chun2010,Chung2013}. However, this method is inefficient in
  measuring the statistical significance of whether the parameters associated with certain
  variables are effectively zero \citep{OlsonHunt2014}. In our implementation, the
  thresholding parameter involved in sPLS is estimated using a 10-fold cross-validation criteria.    

{\bf Sliced Inverse Regression (SIR) with CSS selection}
  In general, SIR methods \citep{Li1991}  reduce the excess dimension in a non-parametric setting through the
  estimation of the linear space spanned by the coefficients of the covariables,
  also known as \textit{effective dimension reduction} (EDR) subspace. 
  This subspace is obtained by an approximate eigenvalue decomposition that
  involves an estimation of the covariance matrices of the design matrix and
  the conditional expectation of the explanatory variables given the
  response. The
  estimation of the covariance matrix requires to partition the dependent variable into subgroups, called \textit{slices}.
  The SIR method can capture both linear or nonlinear associations between the response and
  covariates. However, the estimation of the dimension
  reduction space does not actually lead to a variable selection procedure and the
  covariance estimation relies heavily on the homogeneity of the response within
  each slice \citep{Wu2010}. Because of this, we opt to incorporate the CSS
  (\textit{Closest submodel selection}) variable 
  selection procedure into the SIR method. Furthermore, to better deal with the fact that there is a larger number of
  covariates than observations, we employed the SIR-QZ algorithm, an upgrade of the SIR method based on the generalized Shur decomposition for underdetermined cases \citep{Coudret2014,Coudret2017}.   
  Finally, we studied the association between proxies and temperatures
  through a linear regression between the observed anomalies and a number of EDR directions, i.e. an orthogonal basis of the EDR subspace, determined by marginal dimension tests \citep{Cook2004}. 
  
{\bf Principal Component Regression (PCR)}
PCR simply means that we replace the original covariates by their PCs. To select how many and which PCs to use, we fitted a regression model between the temperature and PCs of eight different sets of covariates selected in each of the eight nests (Table 1) based on the training data from the calibration period. We then selected the number of PCs in each of
the eight regressions as the minimum number that attains, for the first time, an adjusted
$R^2$ of at least 70\% in each case. PCR is often used when covariates are
highly-correlated or when the number of covariates is larger than the number of 
 observations. A caveat of this method is that the principal components
with smaller contribution to variance are not necessarily the ones that
associate less with the dependent variable in a linear regression model 
\citep{Jolliffe1982,Tibshirani1996}. 

{\bf Supervised Principal Components (sPCR)}
Because of the above-mentioned caveat of directly using PCR, \cite{Bair2006} developed a technique where PCR is applied only
to a certain subset of covariates that exhibits a considerable amount of association
 with the dependent variable, and the threshold of ``considerable" is chosen through
cross-validation. Compared to PCR, the sPCR ensures the dimensionality
reduction on the covariates space while taking the association between
the covariates and the dependent variable into account. In general, its performance is quite similar to sPLS  \citep{Chung2013}.  

\begin{figure}[h!]
  \centering
 \includegraphics[scale=0.40]{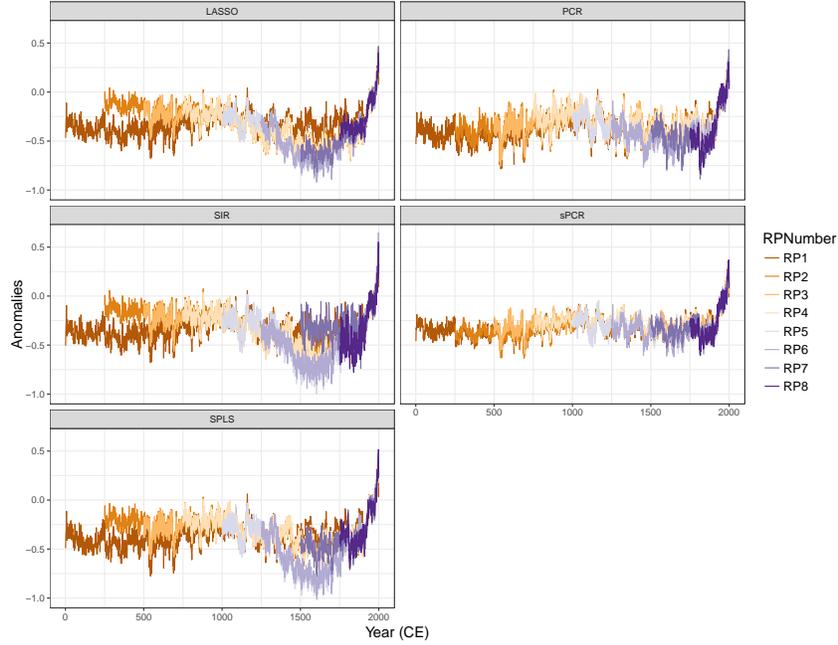} 
  \caption{Reduced Proxies among methods.}
  \label{fig:RPs}
\end{figure}

\begin{figure}[h!]
  \vskip -1cm
  \centering
 \includegraphics[scale=0.35]{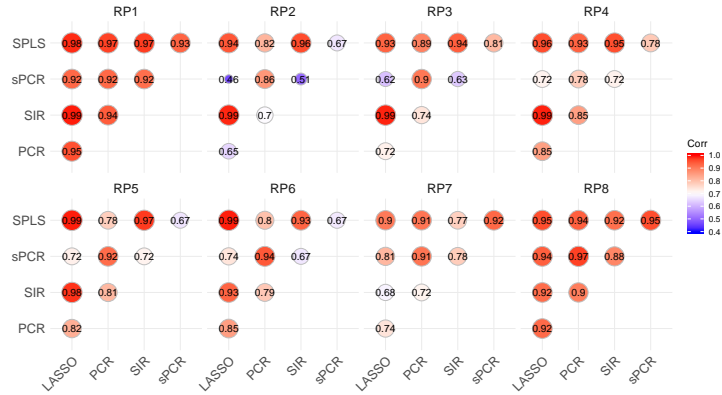} 
  \vskip -1.1cm \caption{Correlation Matrices among 5 different Reduced Proxy series}
  \label{fig:CorrRPs}
\end{figure}

The data reduction allows us to fit linear regression models between
temperatures and proxies. Reducing proxies not only eases the computation, but
also makes the reconstruction less uncertain by removing the part of noise in
proxies that is related to local temperatures. After we fit a linear regression model under each of
the five data reduction methods and for each of the eight proxy groups listed in
Table \ref{tab:distdate}, we compute a single reduced proxy series following
B14 for each group. All reduced proxies are shown in Fig.~\ref{fig:RPs}.  These series are highly correlated, with the reduced proxies obtained by PCR standing out as least similar to the others by this metric (Fig.~\ref{fig:CorrRPs}).

\subsection{Model Specification}
\label{sec:modelspec}
 The first level of a BHM always models the likelihood of the data \citep{Tingley_QSR2012}. The second level models the temperature process, and the third models observational uncertainties. 
 
As in  \citet{boli1} and B14, our process model includes radiative forcings, since they are known to drive the temperature process. This also allows to {\it
  attribute} forcings \citep[i.e., determine causality,][]{HegerlZwiers:2011} as
part of the inference procedure. However, this raises the spectre of
overfitting, whereby the model would discount the noisy proxy data and place
undue weight on the forcings, and not enough on internal variability. To
mitigate this risk, it may be preferable to model temperature fluctuations as a
smooth function of time (say, via splines) without including forcings, then
perform forcing attribution on the inferred temperature posterior
\citep{Schurer2013a, Schurer2013b}, which guarantees independence: this way, if the reconstruction bears a strong resemblance to the forcings, it is only because the latter are reflected in the values of the predictors, not because they were fed to the model. In this section we explore both end-members, as well as an intermediate case. We first define:
\vspace{-.3cm}
\begin{itemize}
\item $RP_t^i$: $i$-th reduced proxy at time $t$.
  
\item $T_t$: temperature anomaly at time $t$.
  
\item $\tilde C_t = \log (C_t)$: Transformed greenhouse gases. The log
  transformation is chosen to approximate the radiative forcing due to changes
  in the equivalent CO$_2$ concentration (see B14).
  
\item $\tilde V_t = \log (-V_t+1)$: Transformed volcanic forcing. See more details in B14 for the choice of the transformation.
  
\item $B_t^{k,\tau}$: $k$-th B-Spline basis function at time $t$ with a uniform knot
  sequence $\tau$ \citep{DeBoor2001,Ramsay2005}. Here we choose
  cubic B-spline bases and we denote $K(\tau)$ as the total number of basis elements.  
\end{itemize}
We then define the first level of BHM as $RP_t^i=\alpha_0^i+\alpha_1^iT_t+\epsilon^i_t$,  
where $\{\alpha^i_j\}$ are intercepts ($j=0$) and slopes ($j=1$) for $i=1,\ldots,N$,  and $\epsilon^i_t$ are normally-distributed random variables with finite variances
$\{\sigma^2_i\}$. Note that in our case $N=8$; the time variable $t$ is defined on each nest according to the
intervals of Table~\ref{tab:distdate}. We explore three formulations of the process level:

\noindent {\bf No forcing (model "NF")}
The main idea of this model is to provide a baseline that ignores external forcings. To capture low-frequency behavior, we include a smoothing function as: 
  \begin{align}\label{eq:M1}
    T_t=\beta_0+\sum_{k=1}^{K(\tau)}\beta_k B_t^{k,\tau}+\eta_t,
  \end{align}
where $\beta_k$ are coefficients for B-spline bases, and
$\eta_t$ are also normally-distributed random variables with finite variances
$\sigma^2_{\eta}$. For simplicity, the error terms $\epsilon^i_t$ and $\eta_t$
are assumed to be independent.  

\noindent {\bf With forcing (model "WF")}
Like B14, this model explicitly models temperature as linearly driven by radiative forcing:
  \begin{align}\label{eq:M2}
    T_t=\beta_0+\beta_1S_t+\beta_2\tilde V_t+\beta_3\tilde C_t+\eta_t.
  \end{align}
Note that Models NF and WF simply assume an IID error structure. This is because B14 
found that complex error structures make little difference when forcings are
added to the reconstruction and if we allow for a more flexible AR(1) structure for the error terms, the conclusions remain the same.

\noindent {\bf "Mixed" Model} 
Finally we consider the more realistic case where temperature reflects both external forcings and internal dynamics. This model is a combination of equations \eqref{eq:M1} and \eqref{eq:M2}, as follows:
  \begin{align}\label{eq:M3}
    T_t=\beta_0+\beta_1S_t+\beta_2\tilde V_t+\beta_3\tilde C_t+\sum_{k=1}^{K(\tau)}\gamma_k B_t^{k,\tau}+\eta_t
  \end{align}
  where $\gamma_k$ are the coefficients for the B-Spline basis. 
  This last case is most relevant to climate dynamicists, as it explicitly represents both endogenous and exogenous variability. The downside is that the model is more complex, thus making estimation more challenging.

\subsection{Computing posteriors with INLA}

The computational challenge of MCMC inference has been a concern for Bayesian paleoclimate reconstructions. It is crucial to overcome this  bottleneck before we can move forward to a more complex space-time reconstruction. Here we introduce the INLA sampling strategy to accelerate the MCMC procedure, as a proof of concept for more comprehensive models. The INLA approach is applicable to a general specification for which the mean $\eta_i$ of the observations $y_i$ follows a linear structure:


\begin{align}\label{eq:meanINLA}
  \eta_i = \alpha +\sum_{m-=1}^M\beta_mx_{mi}+\sum_{l=1}^Lf_l(z_{li})
\end{align}
where $\alpha$ represents an intercept, the coefficients
$\mathbf{\beta} = (\beta_1,\ldots,\beta_M)$ relate $M$ covariates
$(x_1,\ldots,x_M)$ to $\eta_i$, and $f = \{f_1(\cdot),\ldots,f_L(\cdot)\}$ is a collection of
random effects defined on a set of $L$ covariates $(z_1,\ldots,z_L)$ (see
\cite{Rue2009} and \cite{Blangiardo2013}). 
Denote the set of random variables as
$\theta = (\alpha,\beta,f)$ with $K$ hyperparameters $\psi =
\{\psi_1,\ldots,\psi_K\}$, and the vector of observations as $y=(y_1,\ldots,y_n)$. Model \eqref{eq:meanINLA} leads to conditional independence of $y$ given $\theta$ and $\psi$:
\begin{align*}
  p(y|\theta,\psi)=\prod_{i=1}^np(y_i|\theta_i,\psi).
\end{align*}
In our models, if we consider $y_i$ as the reduced proxies,  $\eta_i$ the
linear mean of the reduced proxies, $x_m$ the external forcings and/or the set
of spline basis, and $f(z_l)$ the latent variables (temperature anomalies in
our case), then our models fall into the general specification of INLA. 
The main objectives of our Bayesian estimation are to compute the
marginal posterior distribution of each parameter in $\theta$:
\begin{align*}
  p(\theta_i|y) = \int p(\theta_i,\psi|y)d \psi = \int p(\theta_i|\psi,y)p(\psi|y)d \psi
\end{align*}
To attain computational advantages, INLA assumes that the prior of
vector $\theta$ is a multivariate normal random vector with a precision matrix
that depends on hyperparameters $\psi$. INLA further 
approximates the two components $p(\psi|y)$ and $p(\theta_i|\psi,y)$. The first component is replaced by its a
Laplace Approximation (see \cite{Tierney1986}):
\begin{align*}
  p(\psi|y)=\frac{p(\theta,\psi|y)}{p(\theta|\psi,y)}\propto \frac{p(\psi)p(\theta|\psi)p(y|\theta)}{p(\theta|\psi,y)}
            \approx  \frac{p(\psi)p(\theta|\psi)p(y|\theta)}{\tilde p(\theta|\psi,y)} \Biggm |_{\theta=\theta^*(\psi)} := \tilde p(\psi|y),
\end{align*}
where $\tilde p(\theta|\psi,y)$ is the Gaussian approximation of
$p(\theta|\psi,y)$ and $\theta^*(\psi)$ is its mode (see \cite{Rue2009}). The
second component can be approximated in a similar way:
\begin{equation}\label{eq:second}
  p(\theta_i|\psi,y)  =\frac{p((\theta_i,\theta_{-i})|\psi,y)}{p(\theta_{-i}|\theta_i,\psi,y)} 
   \approx \frac{p((\theta_i,\theta_{-i})|\psi,y)}{\tilde p(\theta_{-i}|\theta_i,\psi,y)} \Biggm |_{\theta_{-i}=\theta_{-i}^*(\theta_i,\psi)}:=\tilde p(\theta_i|\psi,y),
\end{equation}
where $\theta=(\theta_i,\theta_{-i})$, $\tilde p(\theta_{-i}|\theta_i,\psi,y)$ is the Gaussian approximation of
$p(\theta_{-i}|\theta_i,\psi,y)$ and $\theta_{-i}^*(\theta_i,\psi)$ is its mode,  
The approximation in \eqref{eq:second} possesses good precision, but it is very time demanding because it
requires to recompute $\tilde p(\theta_i|\psi,y)$ for each value of $\theta$ and $\psi$. A
more efficient approach is to use the simplified Laplace Approximation that is
based on a Taylor's expansion of $\tilde p(\theta_i|\psi,y)$ in 
\eqref{eq:second}. As mentioned in \cite{Rue2009} and \cite{Blangiardo2013},
INLA first explores the marginal joint posterior of the hyperparameters $\tilde
p(\psi | y)$ in order to locate the mode and then performs a grid search to 
produce a set of ``relevant'' points $\{\psi^*\}$ together with a set of weights
$w_{\psi^*}$ as an approximation of this marginal distribution. The marginals
$p(\psi^*|y)$ are then refined using interpolation methods. Finally, the marginals
$\tilde p(\theta_i|y)$ are obtained as follows:
\begin{align*}
  \tilde p(\theta_i|y) \approx \sum_{\psi^*}\tilde p(\theta_i|\psi^*,y)\tilde p(\psi^*|y)w_{\psi^*}.
\end{align*}

\subsection{Priors}
The models proposed in section \ref{sec:modelspec} were
implemented using the R package \textbf{r-inla} (www.r-inla.org). The
implementation followed the methods provided in \cite{Ruiz-Cardenas2012} and
\cite{Muff2015} on the use of the INLA methodology in state-space models,
dynamic linear models, and in general models whose mean can be written
into equation \eqref{eq:meanINLA}.


Like all Bayesian procedures, INLA requires priors for unknown parameters, given below:
\begin{itemize}
\item $\alpha^i_j\sim N(0,3)$, $\beta_\ell \sim N(0,3)$ for $i=1,\ldots,N$, $j=0,1$, and  $\ell=0,\ldots,3$
  for Model WF and $\ell=0,\ldots,K(\tau)$ for Model NF. The choice of the variance is
  completely arbitrary, but the main idea is to select a relatively large one.
  
\item $\rho_i := -\log \sigma^2_{\epsilon^i}\sim \text{log-gamma}(1,10^{-20})$
  (very small precision) for $i=1,\ldots,N$.
  
\item $\rho_0 := -\log \sigma^2_\eta \sim \text{log-gamma}(1,10^{-20})$ (very
  small precision).
\end{itemize}

\section{Results.}
\label{sec:results}

In the following, the results use model WF for comparison with B14, unless otherwise specified.


\subsection{Impact of data reduction choices}
As a first exercise, we analyze the change in the predictive capacity of the
 model when more equations involving proxies are included. We used
two proper scoring rules \cite{Gneiting2007a} as measures of predictive ability: the Interval Score at $\alpha$ level
(IS$_\alpha$) and the Continuous Ranked
Probability Score (CRPS). These scores have been previously employed in the
verification of point forecasts in environmental sciences for example, as well as the area
of paleoclimatic reconstructions (see B14 and
\cite{Scheuerer2014}). Table~\ref{tab:comparisontot} reports the predictive
measures using INLA's prediction intervals and the observed
  anomalies over 1850-1899 as an out-of-sample validation interval. We used INLA's ability to compute direct
posterior densities when computing interval scores and, in the CRPS case, we used Monte Carlo samples obtained with INLA. 

Validation to the early instrumental record, important though it is, says little about a reconstruction's behavior on centennial scales, which is of primary interest to climate scientists. To constrain this behavior, we leverage the estimates of \citet[hereafter, PS04]{Pollack2004} whose borehole-based
temperature inversions estimate surface temperature trends over 1600-1899. Because of the diffusive transfer of heat in Earth's crust, this dataset only speaks to centennial trends; we thus compare it to smoothed versions of our reconstructions, obtained through a Butterworth low-pass filter with a 100-year cutoff and order equal to 4. We compared in this case Model WF with a single RP (the longest available) with respect to model WF using the 8 available reduced proxies, where the comparison was made under the five dimension reduction methods explained above. The results of this comparison are reported as mean square error (MSE) in the rightmost column of Table~\ref{tab:comparisontot}.

\begin{table}[h!]
  \centering
  {\small
    \begin{tabular}{lll|rrrr|r}
  \toprule
 \textbf{Model} & \textbf{N} & \textbf{Method} & IS$_{80}$ & IS$_{95}$ & CRPS & MSE & MSE(PS04) \\ 
  \midrule
WF & 1 & PCR & 0.4678 & 0.1792 & 0.1641 & 0.0692 & 0.1483 \\ 
WF & 1 & sPCR & 0.5821 & 0.2146 & 0.2593 & 0.1987 & 0.1981 \\ 
WF & 1 & LASSO & 0.4658 & 0.1780 & 0.1655 & 0.0714 & 0.1709 \\WF & 1 & SPLS & 0.4749 & 0.1813 & 0.1810 & 0.0899 & 0.1618 \\ 
      WF & 1 & SIR & 0.4779 & 0.1819 & 0.1671 & 0.0722 & 0.1670 \\
      \midrule
WF & 8 & PCR & 0.2493 & 0.0840 & 0.0986 & 0.0296 & 0.0360 \\ 
WF & 8 & sPCR & 0.1969 & 0.0719 & 0.0745 & 0.0160 & 0.0892 \\ 
WF & 8 & LASSO & 0.5044 & 0.3128 & 0.1559 & 0.0462 & 0.1679 \\WF & 8 & SPLS & 0.3436 & 0.1687 & 0.1177 & 0.0319 & 0.1546 \\ 
      WF & 8 & SIR & 0.6485 & 0.4797 & 0.1871 & 0.0569 & 0.2074 \\
      \midrule
NF & 8 & PCR & 0.2966 & 0.0958 & 0.1249 & 0.0473 & 0.0223 \\ 
      NF & 8 & sPCR & 0.4275 & 0.1624 & 0.1600 & 0.0702 & 0.0237 \\
      NF & 8 & LASSO & 0.5434 & 0.3545 & 0.1644 & 0.0491 & 0.1772\\ 
NF & 8 & SPLS & 0.2617 & 0.0942 & 0.0967 & 0.0248 & 0.1294 \\ 
      NF & 8 & SIR & 0.5772 & 0.3955 & 0.1720 & 0.0516 & 0.1915 \\
      \midrule
Mixed & 8 & PCR & 0.3509 & 0.1101 & 0.1579 & 0.0745 & 0.0157 \\ 
Mixed & 8 & sPCR & 0.3660 & 0.1368 & 0.1357 & 0.0512 & 0.0172 \\ 
Mixed & 8 & LASSO & 0.5291 & 0.3385 & 0.1613 & 0.0480 & 0.1735 \\ 
Mixed & 8 & SPLS & 0.3131 & 0.1404 & 0.1101 & 0.0294 & 0.1463 \\ 
Mixed & 8 & SIR & 0.5986 & 0.4209 & 0.1766 & 0.0533 & 0.1969 \\ 
   \bottomrule
\end{tabular}
}
\caption{Comparison of predictive measures. The IS, CRPS and MSE measures are computed against HadCRUT4 temperature and computed over 1850-1899 as an out-of-sample validation period. The MSE(PS04) pertains to PS04 and is computed over 1600-1899. All measures are negatively oriented, so lower scores reward better estimates.}
\label{tab:comparisontot}
\end{table}

An improvement is evident in all measures when we use all the available reduced proxies over the case
$N=1$, under the methods SPLS, PCR and sPCR for the first validation period.
For the PS04 dataset validation, all the methods except SIR show an improvement in terms of MSE. (Table~\ref{tab:comparisontot}, last column). Also note that, among the models with external forcings, the best performances from the viewpoint of these prediction measures are obtained with sPCR and PCR.

\subsection{Impact of Model Choice}
We are also interested in assessing whether a linear combination of B-spline bases can model GMST without the inclusion of external
forcings at all. It is clear from equations \eqref{eq:M1} and \eqref{eq:M3} that
one of the drawbacks of Models NF and Mixed is the arbitrariness of $K(\tau)$. We analyze the
relationship between the temperature observed during the calibration period
(1900-2000) and a linear combination of BSplines. The number of 
bases in Model NF is selected according to the adjusted $R^2$ of a linear regression model between observed anomalies and
the corresponding basis functions. Based on the above, we selected 6
BSpline functions for this period and we take $K(\tau)=120$ based on the
assumption that the number of BSpline bases is uniform
throughout the entire reconstruction period. Note that the high prevalence of
annually-resolved proxy data allows to assume that a constant number of BSpline bases might be adequate to describe the temperature
mean but we know that the location and spacing of knots can possibly affect the results.  A compelling alternative to our choice is to use a properly penalized spline with a well
defined interpretation in the limit as the number of knots approach infinity. We leave this issue for future investigation..
  
  The choice of $K(\tau)=100$ for model Mixed is
based on the same criteria as before. Finally, we fit the models NF and Mixed
with the previous choices of $K(\tau)$, using the five different data reduction
techniques described in section \ref{sec:rp}. These results are
also shown in Table \ref{tab:comparisontot}. Note that the SPLS method achieves a better performance in terms of the
predictive measures for the first validation period, and PCR gets the smallest
MSE for the second period. 

\begin{figure}[h!]
  \centering
  \includegraphics[scale=0.40]{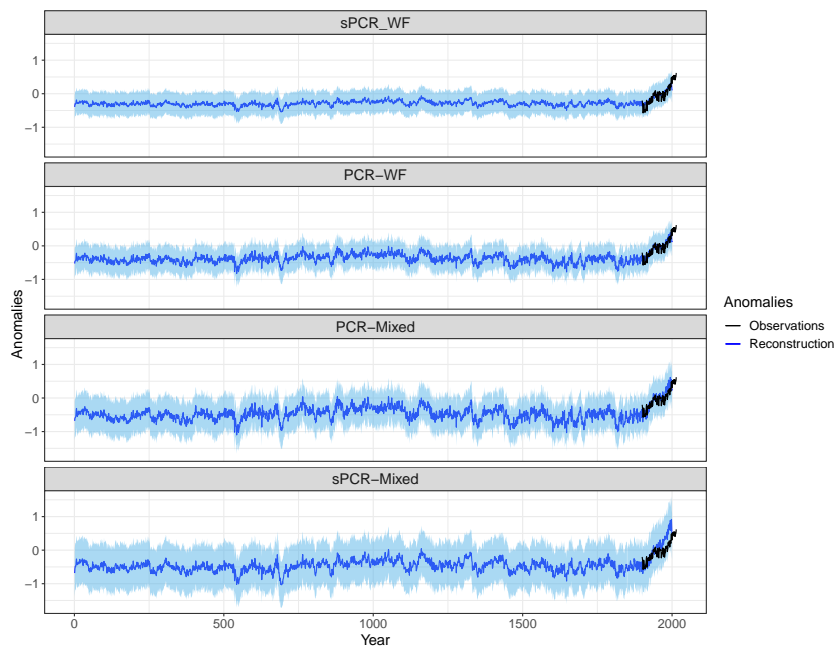}
  \caption{Paleoclimate Reconstruction in the Common Era (CE) with 95\%
    prediction bands. Best two choices per validation method.}
  \label{fig:paleoCE1}
\end{figure}

 Four reconstructions of Common Era GMST
are shown in Figure \ref{fig:paleoCE1}.
In order to illustrate the reconstruction that we obtained for both models WF and Mixed, we considered the best two choices in terms of the validation measures for the
first out-of-sample period  and the best two choices for the second testing
period: sPCR and PCR methods for both the WF and Mixed models.

\begin{figure}[h!]
  \centering
  \subfloat[]{\includegraphics[scale=0.33]{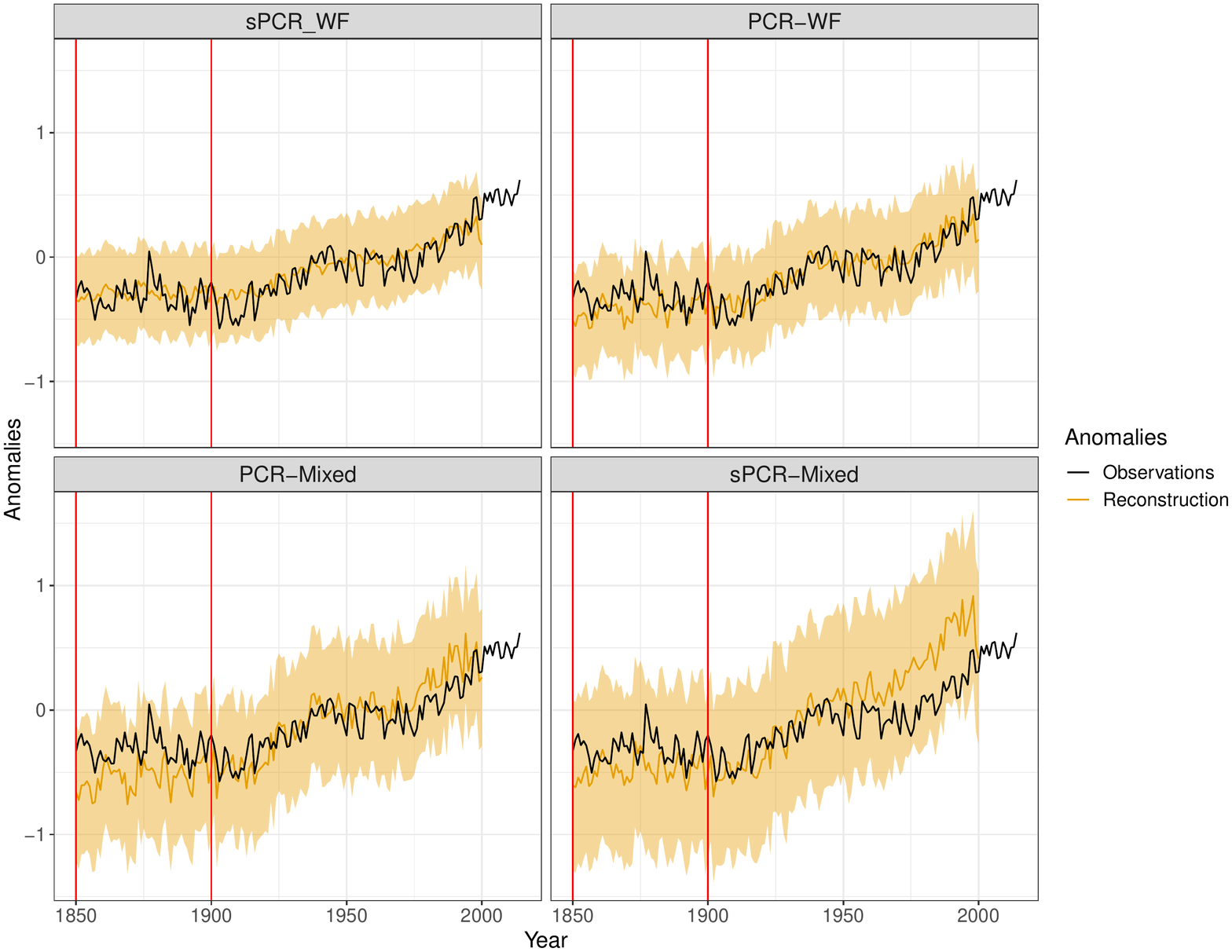}}
  \subfloat[]{\includegraphics[scale=0.33]{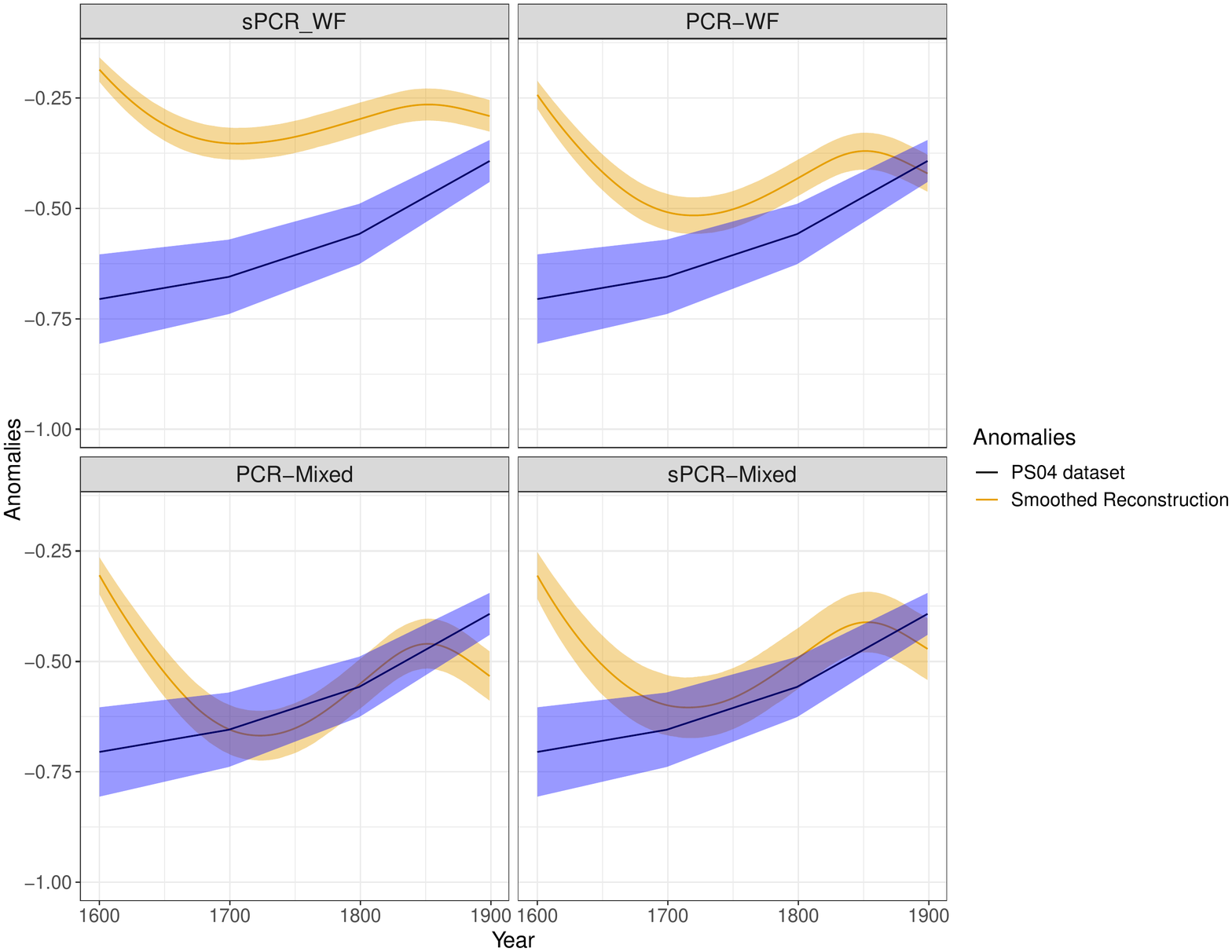}}
  \caption{Panel (A): Paleoclimate Reconstruction 1850-2000 with 95\%
    prediction bands. The out-of-sample validation period is
    located between the red lines. Panel (B): Smoothed Reconstruction 1600-1900
    vs PS04 series with their corresponding 95\% prediction intervals. Best two choices per validation method.}
  \label{fig:paleo19001}
\end{figure}

The best model for the first testing period (sPCR-WF) shows an interesting
balance in terms of the variance of the reconstructed series and the width of
the confidence region approximated by INLA. The remaining 3 reconstructions show
similar small-scale tendencies with respect to the best choice, but the width of
their confidence regions is greater. A closer look of the reconstructions is shown in Figure
\ref{fig:paleo19001}, Panel (A), where the first testing period appears between the red lines.
Note that sPCR-WF closely predicts the anomalies observed in the first testing
period and it does so with a higher level of accuracy than the other
reconstructions. In addition, the other three reconstructions underestimate in
an almost analogous way the anomalies observed during the same period. At least
for this validation exercise there is no clear advantage to using a B-Spline
basis as an additional linear term in equation \eqref{eq:M3}.

Figure \ref{fig:paleo19001}, Panel (B) shows a comparison of the smoothed reconstructions
using a low-pass filter and the borehole-based reconstruction of
\cite{Pollack2004} (PS04 dataset). In this case, the mixed versions of the sPCR
and PCR models with external forcings provide the best adjustment in terms of
the MSE measure, especially during the 1700-1850 period. We can infer therefore
that the B-Spline basis provides enough flexibility to attain reconstructed
anomalies with low-frequency features. Therefore, PCR-Mixed yields the most balanced
results in terms of these two validation exercises.


\subsection{Impact of INLA sampling}

\begin{figure}[h!]
  \centering
  \includegraphics[scale=0.40]{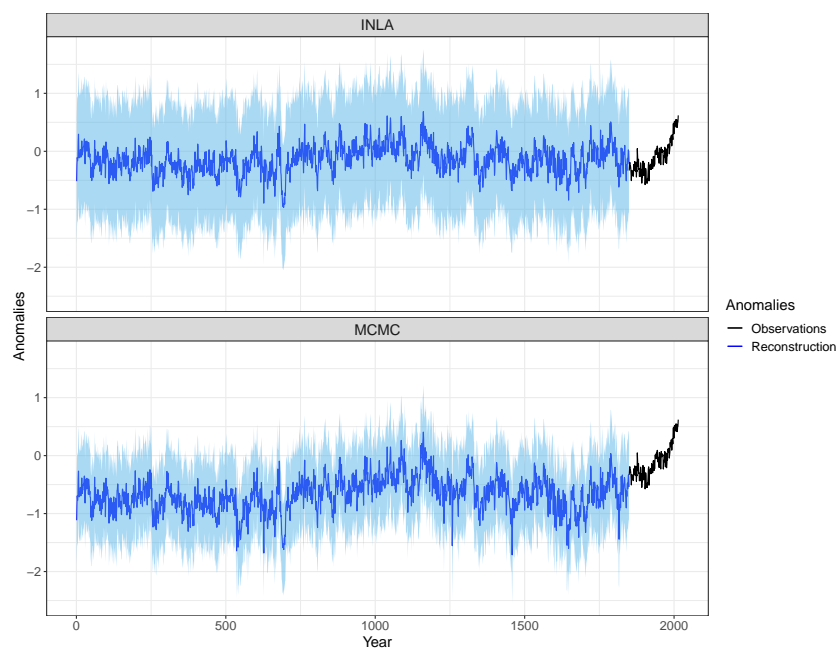}
  \caption{Paleoclimate Reconstruction in the Common Era (CE) with 95\%
    prediction bands. Model sPCR-WF under two methods: INLA and MCMC.}
  \label{fig:paleoCE4}
\end{figure}

\begin{figure}
  \centering
  \includegraphics[scale=0.3]{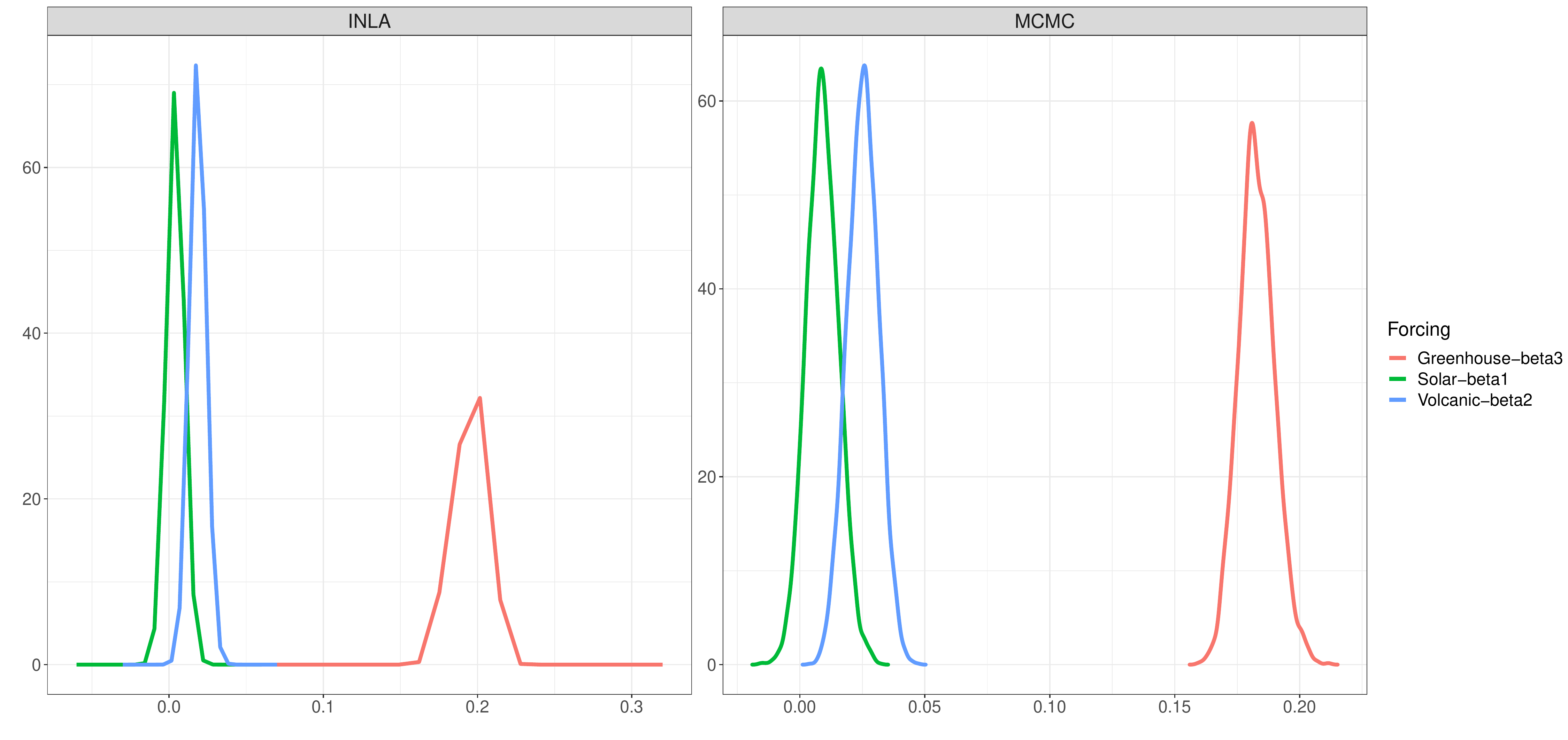}
  \caption{Posterior densities of $\beta_1$, $\beta_2$ and $\beta_3$ for Model
    sPCR-WF and MCMC using PCR single reduced proxy.}
  \label{fig:betas}
\end{figure}

We now quantify the trade-offs of approximating the MCMC procedure using INLA. The WF model in its simplest case (1 nest) was fitted in B14
using an MCMC approach. We employed this approach in order to adjust the
WF model with the first reduced proxy from the sPCR method (chosen purely
for comparison purposes). The MCMC was performed using the
same priors as in B14, but with a larger calibration period
(1900-2000). The results are shown in Figures~\ref{fig:paleoCE4} and
\ref{fig:betas}. Note that the MCMC reconstruction reaches cooler temperatures
than INLA's and that its confidence bands along the reconstruction period are
narrower. This last fact coincides with a small difference in terms of the
interval score measure. Despite these contrasts, the general trends of both reconstructions are quite similar. 

Another point of comparison is the estimated coefficients of the
external forcings in equation \eqref{eq:M2}. The estimated density function
for each coefficient is shown in Figure \ref{fig:betas}. The behavior of the
estimated parameters of the three external forcings is very similar between the
two methods: by far the most influential external forcing in
both reconstructions is the greenhouse gas component, followed by explosive volcanism. Consistent with \citet{Schurer2013b}, solar irradiance comes last, and is indistinguishable from a zero effect. 

For the same reason as above, the estimated density
function of the CO$_2$ coefficients are more
concentrated when we use an MCMC algorithm, to a lesser extent for the other 
coefficients. This behavior confirms the observed details of Figure
\ref{fig:paleoCE4}. 

In terms of computational efficiency, the INLA procedure is quite remarkable, not only
for our case but in most of the previous work on a similar topic as well (see \cite{Rue2009},
\cite{Blangiardo2013}, \cite{Ruiz-Cardenas2012} for a few examples). The
computational cost of MCMC sampling with 5000 samples was
approximately 8 hours, whereas the computational time of INLA's best model with
a single reduced proxy was approximately 15 seconds. This comparison
was performed on an Ubuntu 16.04 server with Intel Xeon E5-2630 (8-cores,
2.40GHz) and 64 GB of RAM.  This large speedup allowed us to explore a far wider variety of modeling choices than MCMC alone.

\section{Conclusions}
We carried out a Bayesian inference of global mean surface temperature over
the past 2,000 years. By leveraging INLA to lighten the computational burden, our framework allows us to investigate a wide range of model choices and data reduction strategies.  We validated the result using instrumental data over the 1850-1899 period, and using independent borehole temperature inversions over 1600-1899. The former validates the reconstruction precisely to an interannual scale, while the latter constrains centennial trends.  

Below are a few take-home messages:\\
     (a) The data reduction techniques provide roughly equivalent results, with
      sPCR and PCR performing marginally better than other methods.
    
     \noindent (b) The model choice is highly consequential. Model "Mixed" is the most
      physically justifiable, and it guarantees a balance between the
      validation measures on the first and second testing periods. This models also
      appears to perform at least as well as other choices. The 
      additional nonparametric terms in the mean component of equation \eqref{eq:M3} allows
      to capture long-memory behavior that is not included in the external forcings, which compensates for the independent structure of the errors. 
      
     \noindent (c) In cases where both INLA and MCMC are implemented, INLA allows to approximate the MCMC
      solution at a fraction of the computational cost, but with wider
      prediction intervals. This exercise was performed only for the simplest
      model available (1 reduced proxy), because of the prohibitive cost of the
      MCMC approach for more complex models like the ones presented in Figure
      \ref{fig:paleoCE1}. 
      
     \noindent (d) Adding to a wide body of literature, we find that current temperature levels are unprecedented in the past 2,000 years. Both the models that include forcings (WF, Mixed) show that the man-made increase in atmospheric carbon dioxide is the leading contributor to this warming effect. 

One limitation of our analysis is the strong
dependence of our results on the choice of the testing period. Ideally, a
cross-validation exercise should be carried out to determine with greater
certainty the expected prediction error of the models, but due to the restricted access to
additional comparison information the cross-validation is very challenging for this problem. 
Another way to improve our current study is to consider the space-time variability of the reduced proxies and temperatures in the model. 
Since INLA has been proven to be computationally
efficient, it could be used to extend this work to a spatiotemporal context. Yet another important area of application is the simultaneous inference of climate fluctuations and deposition age \citep{Sweeney:WIRES2018}.

\footnotesize

\end{document}